\RequirePackage{lineno}

\documentclass[aps,prc,twocolumn,amsmath,amssymb,groupedaddress,superscriptaddress,nofootinbib]{revtex4-2}

\usepackage{graphicx}
\usepackage{bm}

\usepackage{xcolor}
\usepackage{lineno}
\usepackage{soul}

\begin{document}

\title{Detailed Study of Quark-Hadron Duality in Spin Structure Functions of the Proton and Neutron}

\author{V. Lagerquist} 
\affiliation{Old Dominion University, Norfolk, Virginia 23529, USA}
\author{S.E. Kuhn}
\email[Contact author. Email: ]{skuhn@odu.edu}
\affiliation{Old Dominion University, Norfolk, Virginia 23529, USA}
\author{N. Sato} 
\affiliation{Thomas Jefferson National Accelerator Facility, Newport News, Virginia 23606, USA}

\begin{abstract}
\begin{description}

\item[Background] The response of hadrons, the bound states of the strong force (QCD), to external probes can be described in two different, complementary frameworks: As direct interactions with their fundamental constituents, quarks and gluons, or alternatively as elastic or inelastic coherent scattering that leaves the hadrons in their ground state or in one of their excited (resonance) states. The former picture emerges most clearly in hard processes with high momentum transfer, where the hadron response can be described by the perturbative expansion of QCD, while at lower energy and momentum transfers, the resonant excitations of the hadrons dominate the  cross section. The overlap region between these two pictures, where both yield similar predictions, is referred to as quark-hadron duality and has been extensively studied in reactions involving unpolarized hadrons. Some limited information on this phenomenon also exists for polarized protons, deuterons and $^3$He nuclei, but not yet
for the neutron.

\item[Purpose] In this paper, we present 
comprehensive and detailed results on the correspondence between the extrapolated deep inelastic structure function $g_1$ of both the proton and the neutron with the same quantity measured in the nucleon resonance region. 
{
Thanks to the fine binning and high precision of our data, and using a well-controlled pQCD fit for the partonic prediction, we can make quantitative statements about the kinematic range of applicability of both local and global duality.}

\item[Method] 
We use the most updated QCD global analysis results at high-$x$ from the JAM collaboration 
to extrapolate the spin structure function $g_1$
into the nucleon resonance region and then integrate over various intervals in the scaling variable $x$. We compare the results with the large data set collected in the quark-hadron transition region by the CLAS collaboration
{
including, for the first time, deconvoluted neutron data}, integrated over the same intervals. We present this comparison as a function of the momentum transfer $Q^2$.

\item[Results] We find that, depending on the integration interval and the minimum momentum transfer chosen, a clear transition to quark-hadron duality can be observed in both nucleon species. Furthermore, we show, for the first time, the 
approach to scaling behavior for $g_1$ measured in the resonance region at sufficiently  high momentum transfer. 

\item[Conclusions] Our results can be used to quantify the deviations from the applicability of pQCD for data taken at moderate energies, and help with extraction of quark distribution functions from such data.
\end{description}
\end{abstract}

\preprint{JLAB-THY-22-3610}
\maketitle

\section{Introduction}

Quantum Chromo-Dynamics (QCD) is the fundamental theory describing the interactions between quarks and gluons (partons), leading to their observed bound states (hadrons) and the strong nuclear force. At high spatial resolution (momentum scale), the QCD coupling constant becomes small (asymptotic freedom~\cite{Gross:1973id,Politzer:1973fx}), and quark and gluon interactions can be calculated perturbatively (pQCD). This leads to the emergence of these partons as effective degrees of freedom in the description of hard processes like deep inelastic scattering where the observed cross section can be described approximately as an incoherent sum of scattering cross sections on individual point-like and structureless partons. On the other hand, at low momenta and long distance scales, the interaction becomes strong and a perturbative treatment is no longer possible. Instead, physical processes can be best described in terms of effective hadronic degrees of freedom, {\it e.g.}, the excitation of resonant hadronic states. By varying the resolution of a probe from short to long distances, physical cross sections displays a transition from the partonic to the hadronic domains. It remains an important question whether there is a region where both pictures apply simultaneously, {\it i.e.}, whether a parton-based description can reproduce the data in the  kinematic region of hadronic resonances, at least on average.  This phenomenon is  known as Quark-Hadron Duality~\cite{Bloom:1970xb, Bloom:1971ye, DeRujula:1976baf, Melnitchouk:2005zr}.  While strong evidence for duality has been found, it is important to fully test the applicability of this concept in the case where spin degrees of freedom are present, and for different hadronic systems. If quark-hadron duality can be firmly established and its applicability quantitatively described, one can use  measurements of hadronic observables to improve contraints on the parton structure of these hadrons. For instance, measurements of nucleon structure functions that are sensitive to high parton momentum fraction $x$ are very difficult at high energies, which limits our knowledge of the very important behavior of  the underlying Parton Distribution Functions (PDFs) as $x \rightarrow 1$.  If the requirement of avoiding the region of nucleon resonances can be relaxed in a controlled manner,  data taken at lower energies could contribute invaluable information on this asymptotic behavior. 

In the present paper, we present new results on tests of duality in proton and neutron spin structure functions. Following this introduction, we introduce the relevant formalism and theoretical concepts, describe the data set we analyzed as well as the phenomenologically extracted spin structure functions from the JAM QCD global analysis to which we compare these data,
and then present results and conclusions.

\section{Theoretical background}
 
\begin{figure}[htb]
\centering
\includegraphics[width=6cm]{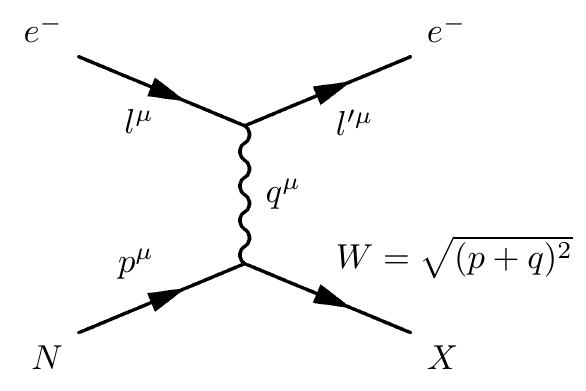}
\caption{Feynman Diagram for inclusive electron scattering off a nucleon target. $W$ is the invariant mass of
the unobserved final state $X$. All other symbols are explained in the text.}
\label{fig:fey1}
\end{figure}

 In this paper, we focus on quark-hadron duality in polarized inclusive electron scattering off polarized nucleon targets. In the single photon exchange approximation, an electron with four momentum $l$ scatters with final momentum $l'$ 
 from a nucleon with momentum $p$ by exchanging a space-like virtual photon with momentum $q=l-l'$ (see Fig. \ref{fig:fey1}). 
 
The invariant cross sections can then be written as \cite{Collins:2011zzd}
\begin{align}
    E'\frac{d\sigma}{d^3 l'} = \frac{2\alpha^2}{s Q^4}L_{\mu\nu} W^{\mu\nu}
\end{align}
where $Q^2=-q^2$ is the virtuality of the exchanged photon, $L_{\mu\nu}$ is the leptonic tensor and $W_{\mu\nu}$ is the hadronic tensor. The latter can be written as a linear combination of unpolarized structure functions $F_{2,L}$  and the polarized structure functions $g_{1,2}$. The polarized structure functions can be experimentally accessed 
by measuring cross sections differences of the form 
\begin{align}
    d\sigma^{\downarrow \Uparrow}
   -d\sigma^{\uparrow \Uparrow}
\end{align}
where ${\downarrow \Uparrow}$ and ${\uparrow \Uparrow}$ corresponds to anti-parallel and parallel beam and target spin configurations, respectively.

In the kinematics of moderate $x= Q^2/2P\cdot q$ and $Q^2$  much larger than hadronic mass scales, the $g_1$ structure function can be approximated in collinear factorization schematically  as 
\begin{align}
g_1(x,Q^2)=& \sum_i \int_x^1 \frac{d\xi}{\xi} \Delta f_{i/N}(\xi,Q^2) 
            \Delta H_i\left(\frac{x}{\xi},\alpha_S(Q^2)\right)\notag\\
            &+O\left(\frac{m}{Q}\right)\; .
\label{e.factorization}
\end{align}
Here the sum runs over all parton flavors $i$. The term $\Delta H_i$ is the target-independent short-distance partonic coefficient function calculable in pQCD in powers of the strong coupling $\alpha_S$ and is convoluted with  the spin-dependent 
Parton Distribution Function (PDF) $\Delta f$ in the variable $\xi$. The factorization theorem is valid up to corrections of the order $m/Q$ where $m$ is a generic hadronic mass scale. The $\xi$ variable is the light-cone momentum fraction of partons relative to the parent hadron,
{\it  i.e.}, $\xi=k^+/p^+$. At leading order in pQCD, the hard factor $\Delta H_i$ is proportional to $\delta(x-\xi)$; hence the structure function $g_1$ has a leading order sensitivity to PDFs at $\xi=x$. Beyond the leading order however, 
the physical structure function receives PDF contributions in the range $x<\xi<1$ due to the convolution in 
Eq.~\ref{e.factorization}.
The scale dependence on $Q^2$ in $\Delta f$ is governed by the DGLAP evolution equations stemming from the renormalization of parton densities and  are  given as 
\begin{align}
    \frac{d\Delta f_i}{d\ln \mu^2}(\xi,\mu^2) = \sum_j \int_{\xi}^1 \frac{dy}{y} 
        \Delta P_{ij}\left(\frac{\xi}{y},\alpha_S(\mu^2)\right) \Delta f_j(y,\mu^2)
\end{align}
where $\Delta P_{ij}$ are the Altarelli–Parisi space-like splitting functions. Finally we remark that the  structure function $g_2$ has no leading power contribution. 

Since the focus of our study is the behavior of $g_1$ in the large-$x$, moderate $Q^2$ regime, it is important to utilize a QCD global analysis framework that has a maximal kinematical overlap in $x$ to allow us to study duality with minimal extrapolation. {
In \cite{Sato:2016tuz}, the Jefferson Lab Angular Momentum Collaboration (JAM)  carried out a comprehensive analysis of the double spin asymmetries in DIS with an extend kinematic coverage in $x$ and $Q^2$ including data with final state mass as low as $W^2=4~{\rm GeV^2}$. To our knowledge, this is the only global analysis that has included systematically all the high-$x$ data from CLAS 6 GeV  with dedicated treatments for twist-3 effects and target mass corrections for the double spin asymmetries.}
In the following, we utilize the inferred $g_1$ from the JAM global analysis (which has a kinematic convergence up to $x\sim 0.7$) and use DGLAP backward evolution to access the resonance region at high-$x$ and lower $Q^2$. 

\begin{figure}[htb]
\centering
\includegraphics[width=8cm]{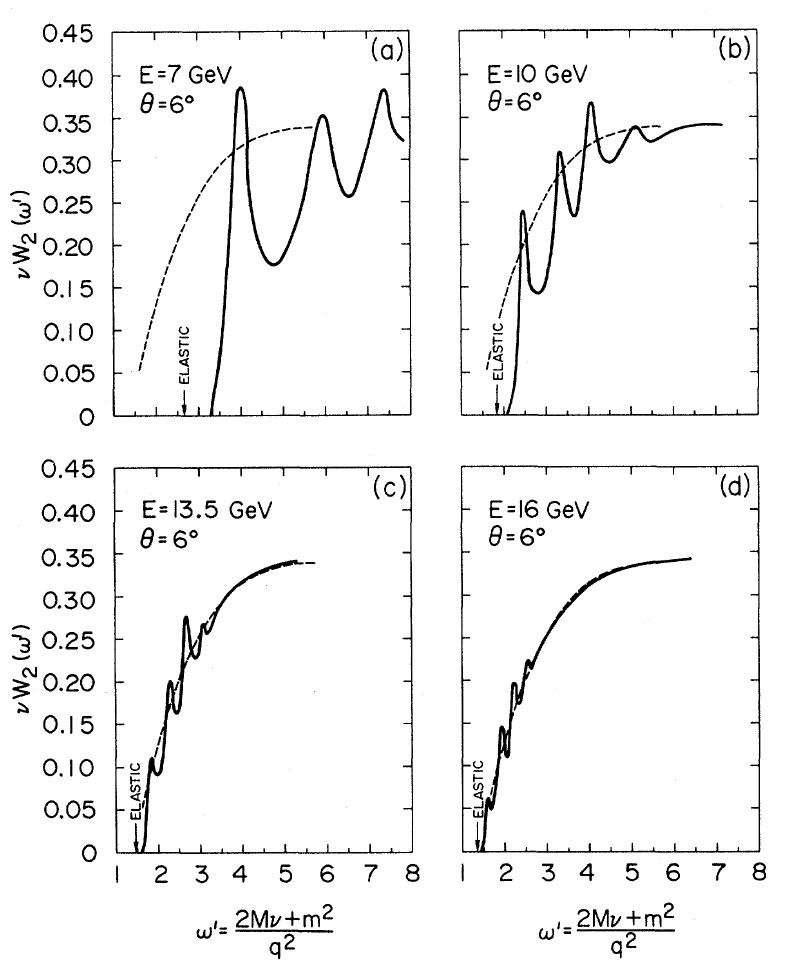}
    \caption{
%
    Schematic dependence of the measured structure function $F_2$ in inelastic electron scattering off the nucleon on the variable $\omega^\prime = W^2/Q^2 +1$, which is close to $1/x$ at large $Q^2$. Panels (a) through (d) are for increasing four-momentum transfer $Q^2$. As can be observed, the resonance excitations of the nucleon are most prominent at low $Q^2$, while at higher $Q^2$ the curve for $F_2$ approaches the scaling limit (dashed line), hence indicating a transition to quark-hadron duality in this observable. Reproduced from the paper by E. D. Bloom and F. J. Gilman~\cite{Bloom:1970xb}, with the permission of AIP Publishing.}
\label{fig:BloomGil}
\end{figure}

For moderate final hadronic state masses, $W<2$~GeV, 
the cross section typically exhibits multiple resonance
peaks that appear when the target is excited into other baryonic states before later decaying into final state products.
This is illustrated in Fig.~\ref{fig:BloomGil} for the $F_2$ structure function. 
This so-called resonance region can be best described in terms of hadronic degrees of freedom, where the cross section is expressed  in terms of transition strengths to the various nucleon resonances,  together with non-resonant hadron production contributions~\cite{HillerBlin:2019jgp}.

It is not a priori obvious how this resonant behavior is related to the underlying degrees of freedom of all hadrons, quarks and gluons, and their description in terms of PDFs, perhaps augmented by higher-twist terms in the OPE. This is addressed by the concept of Quark-Hadron Duality that was first introduced in a publication by Bloom and Gilman in 1970~\cite{Bloom:1970xb,Bloom:1971ye}.  They found that the $F_2$ structure function measured in the nucleon resonance region  approaches a smooth ``scaling curve'' as $Q^2$ increases, with the resonant troughs and peaks approximately averaging out to match an extrapolation of the deep inelastic  structure function at high W into the resonance region (see Fig.~\ref{fig:BloomGil}). 

In particular, Bloom and Gilman proposed that integrals over specific ranges in  $\omega^\prime = 1/x + M^2/Q^2$ (or just over $x$) of either the extrapolated DIS fits or the experimental data in the resonance region would give similar results. The case where the limits of integration cover only 100-200 MeV on either side of a single resonance peak is referred to as local duality, as opposed to global duality which covers the entire resonance region from threshold to $W = 2$ GeV, potentially
also including the elastic peak. In either case, the relation can be summarized as
\begin{align}
 \int_{x_1(W_1,Q^2)}^{x_2(W_2,Q^2)}dx\ F_2^{res}(x,Q^2)
=\int_{x_1}^{x_2}dx\ F_2^{DIS}(x,Q^2),
\end{align}
where $F_2^{res}$ is the structure function { measured} {
over some kinematic range within the resonance region, between $W_2$ and $W_1$ (both below $W = 2$ GeV)}, while $F_2^{DIS}$ is 
extrapolated from a QCD global analysis. 
Here,
\begin{align}
x(W,Q^2) = \frac{Q^2}{W^2-M^2+Q^2} .
\label{xofW}
\end{align}

Since the initial discovery by Bloom and Gilman in 1970, considerable progress has been made in the measurement of unpolarized structure functions at low to moderate $Q^2$ and $W$ and their interpretation in terms of quark-hadron duality, notably at the Thomas Jefferson National Accelerator Facility (also known as Jefferson Lab) \cite{Niculescu:2000tk, Niculescu:2000tj,CLAS:2003iiq, Niculescu:2015wka, Christy:2007ve, Tvaskis:2016uxm, JeffersonLabE00-115:2009jll, Malace:2009dg}. 

In addition to this, spin dependent structure functions  in the same kinematic region have also been studied. Experiments at SLAC in the late 70's provided the first resonance region measurements for polarized proton-electron scattering \cite{Baum:1980mh, Baum:1983ha}. These experiments hinted at the applicability of Bloom-Gilman Duality to proton spin structure functions. They were followed in the 
90's by further experiments at SLAC by the E143 collaboration, which expanded their $g_1$ and $g_2$ measurements to the resonance region \cite{Anthony:1993uf,Abe:1998wq}.
In 2003, the HERMES Collaboration, \cite{Airapetian:2002rw}) published their results specifically on the quark-hadron duality 
for the proton asymmetry $A_1$ measured at 5 different points in average $x$ (corresponding to 5 different regions in average $Q^2$). These data,
together with the E143 ones, were analyzed in a paper by Bianchi {\em et al.}~\cite{BFL} which contrasted, for the first time,
 the kinematic range
where duality seemed to hold  for unpolarized {\em vs.} polarized structure functions. They were followed by  data from
Jefferson Lab (Hall B, \cite{CLAS:2002knl} and Hall A, \cite{Meziani:2004ne}) which contributed to the investigation of spin structure functions in the resonance region with increased kinematical coverage. {
Most of these early experiments had limited statistical
precision and fairly few and wide bins.
The new century brought additional high-precision experiments at Jefferson Lab (Halls A \cite{Zheng:2004ce, Zheng:2003un,JeffersonLabE01-012:2008kgk,Parno:2014xzb, Flay:2016wie}, B \cite{CLAS:2006ozz, CLAS:2008xos, CLAS:2006hmk, CLAS:2015otq, Fersch:2017qrq, Fersch:2018ydg} and C \cite{Wesselmann:2006mw}). 
Even for those, the neutron structure functions in the resonance region were extracted
from measurements on 
nuclei without unfolding their smearing through
nuclear Fermi motion. The present paper uses the most extensive data set available so far, and for the first time includes 
unfolded spin structure functions of the neutron.

Studying quark-hadron duality in the spin sector is important, since polarization dependent observables can have both positive and negative sign, and hence offer a more stringent test of duality}
{
For instance,
it is well-known that the transition to Delta-baryons in the final state are dominated
by the $M1$ amplitude, which should lead to a negative asymmetry $A_1$ and negative $g_1$.
Meanwhile, given the rather low $W$ of the lowest-lying $\Delta(1232)$, the
extrapolated values for $g_1$ from pQCD fits will be at high values of $x$, where
most DIS data indicate positive values for $g_1$}.

In the present paper, we are presenting a new comparison of the most comprehensive data set on spin structure functions in the transition region between hadronic and partonic degrees of freedom, from the EG1b experiment~\cite{Fersch:2017qrq,CLAS:2015otq}, to the recent JAM QCD global analysis~\cite{Sato:2016tuz} at high $x$. 
For the first time, we include the unfolded neutron structure
function $g_1$ in this comparison.
We address both the 
question under what circumstances global and/or local duality holds, and data from
which kinematic region may be used to further constrain pQCD fits without introducing
excessive higher twist corrections.

\section{Input data}

For a detailed study of duality, one needs a dense set of data that cover the entire resonance region (conventionally from $W = 1.072$ GeV to 2 GeV) in fine $W$ bins, for a large number of bins in $Q^2$.  The most comprehensive such data set was collected by the ``EG1b'' experiment carried out with CLAS \cite{CLAS:2003umf} at Jefferson Lab during 2000-2001 \cite{CLAS:2006ozz, CLAS:2008xos, CLAS:2006hmk, CLAS:2015otq,Fersch:2017qrq}. The experiment used the polarized electron beam from the  Continuous wave Electron Beam Accelerator Facility (CEBAF) at Jefferson Lab,  with beam energies of 1.6, 2.5, 4.2, and 5.7 GeV. Together with the large acceptance of CLAS, this set of beam energies yielded a large kinematic reach (with partially overlapping regions), covering nearly 2 orders of magnitude in $Q^2$ ($Q^2 = 0.06...5$) and $W$ from threshold to about 3 GeV.  A particular advantage of the wide acceptance of CLAS is that the data could be sorted into a pre-determined grid of $Q^2$ and $W$, with no need to interpolate between different data points. 
 
The polarized nucleon targets were provided in the form of irradiated frozen ammonia and deuterated ammonia for measurements of proton and deuteron asymmetries, respectively. The target was polarized through Dynamic Nuclear Polarization and reached a polarization along the beam
direction of approximately 75\% for the protons and 30\% for the deuterons \cite{Keith:2003ca}.

The measured double-spin asymmetries were converted into spin
structure functions $g_1(W,Q^2)$ using a phenomenological fit to
the world data on polarized and unpolarized structure functions. In the
case of the neutron structure function $g_1^n$, a folding prescription \cite{Kahn:2008nq} was used to relate the measured spin structure function of the deuteron to $g_1^n$ for each kinematic point. This yielded the first data set of un-integrated neutron spin structure functions in the resonance region. Details about the experiment, the data analysis and the complete data sets can be found in \cite{CLAS:2015otq,Fersch:2017qrq}.

Extrapolated pQCD predictions for $g_1^p$ and $g_1^n$, which are compared to the resonance region data in this paper, are taken from 
the JAM15 fits \cite{Sato:2016tuz} of the world data on inclusive spin observables, including the EG1b data {\em outside} the resonance region ({\it i.e.}, for $W > 2$ GeV). The JAM fits used a novel iterative Monte Carlo fitting method that utilizes data resampling techniques and cross-validation for a robust determination of the uncertainty band of the fitted PDFs as well as any observables predicted from the fit. A total of 2515 data points from 35 experiments and 4 facilities (CERN, SLAC, DESY, and JLab) were included in the fit.

\section{Analysis}

\begin{table}[h!]
\centering
\begin{tabular}{|l| c| c|} 
 \hline
\  & Lower $W$ limit & Upper $W$ limit\\
 \hline
1 & 1.072 & 1.38\\ 
2 & 1.38 & 1.58\\
3 & 1.58 & 1.82\\
4 & 1.82 & 2\\
5 & 1.072 & 2\\ 
6 & 0.939 & 2\\
 \hline
\end{tabular}
\caption{Selected $W$ Ranges, in GeV}
\label{table:W}
\end{table}

In this paper, we investigate two different but related tests of duality:
i) a direct comparison between truncated integrals over measured spin structure functions,
each covering a specific range in final-state mass $W$,
and corresponding integrals over the extrapolated pQCD fits,
 and ii)
a study of the approach to scaling for $g_1$ averaged over a set of fixed narrow ranges in $x$. 

%
{

For the first test, we select six different ranges of $W$ as shown in table \ref{table:W}. 
The first four of these ranges 
cover specific prominent resonance peaks visible
in inclusive unpolarized cross section data (see Fig.~\ref{fig:BloomGil}):
the Delta resonance ($\Delta(1232)3/2^+$), the region of the N$(1440)1/2^+$, N$(1520)3/2^-$, and 
N$(1535)1/2^-$
resonances, the region of the N$(1680)5/2^+$ and nearby resonances, and the remaining region
up to $W = 2$ GeV which does not exhibit a strong peak in the inclusive spin-averaged 
cross section but is known to contain several Delta-resonances.
We test whether local
duality holds in each of these individual resonance regions.
The next range (line 5 in table \ref{table:W}) covers the entire ``canonical'' resonance region,
$1.072$~GeV~$< W < 2$~GeV, {\em i.e.}, the previous four regions combined.
For the last range we add the elastic peak at
$W = 0.939$~GeV, the nucleon mass, to cover the entire region 
$0.939$ GeV $< W < 2$ GeV, extending the corresponding $x$-range up to $x =1$. 
This elastic contribution comes in the form
\[g_1^{el} = \frac{1}{2}\frac{G_EG_M+\tau G^2_M}{1+\tau}\delta (x-1) ,\]
where $G_M = F_1 + F_2$ and $G_E = F_1 - \frac{Q^2}{4M^2}F_2$ are the magnetic and electric Sachs form factors \cite{Walker:1993vj}. The corresponding 
integrals for the JAM extrapolation are simply integrated up to $x = 1$
but don't contain an elastic contribution since they are based on DIS pQCD fits.}

\begin{figure}[htb!]
\centering
\includegraphics[width=0.4\textwidth]{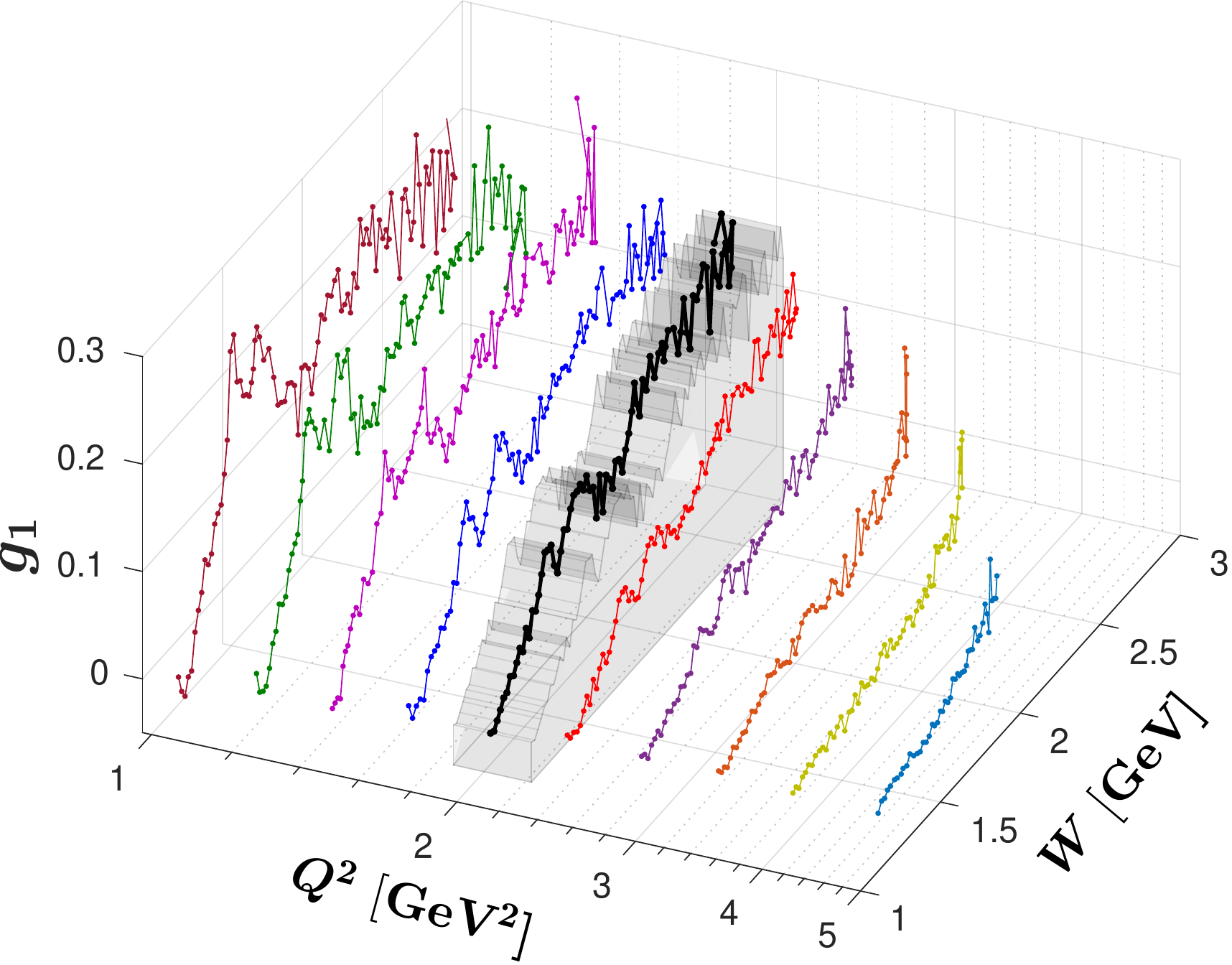}
\caption{Representation of the experimental data set used in this analysis. The measured data points are binned
in bins in $Q^2$, as indicated by the shaded area for the example of the bin 1.87 GeV$^2 < Q^2 <$2.23 GeV$^2$;
see also Table~\ref{table:Q}. The truncated integrals are then formed over specific regions in $W$ as spelled out
in Table~\ref{table:W}}.
\label{fig:diagram}
\end{figure}

\begin{table}[h!]	
\centering
\begin{tabular}{|c|c|c|}						
\hline						
Lower	$Q^2$	&	Upper	$Q^2$	&	Central $Q^2$\\
\hline		
0.92		&	1.10		&	1.00\\
1.10		&	1.31		&	1.20\\
1.31		&	1.56		&	1.43\\
1.56		&	1.87		&	1.71\\
1.87		&	2.23		&	2.04\\
2.23		&	2.66		&	2.43\\
2.66		&	3.17		&	2.91\\
3.17		&	3.79		&	3.47\\
3.79		&	4.52		&	4.14\\
4.52		&	5.40		&	4.94\\
5.40		&	6.45		&	5.90\\
\hline
\end{tabular}	
\caption{Experimental $Q^2$ Ranges, in GeV$^2$}
\label{table:Q}
\end{table}		

For each of these $W$ ranges, our analysis process is the same. Experimental data
for $g_1(x,Q^2)$ are first sorted into bins of $Q^2$ with limits shown in Table \ref{table:Q},
see Fig.~\ref{fig:diagram}.
The $W$-limits for each range are mapped 
to the corresponding 
values for $x$, following Eq.~\ref{xofW}.
The data are then integrated over the corresponding $x$-ranges to yield
the truncated first moments of $g_1$,
\begin{equation}
 \bar{\Gamma}_1(\Delta W, Q^2) = \int_{x_1(W_1,Q^2)}^{x_2(W_2,Q^2)}dx\ g_1(x,Q^2).
\end{equation}

{
The corresponding truncated DIS integral have been calculated by extrapolating the PDF fits of the JAM collaboration to the central $Q^2$ value of each bin.}


For the experimental data, the statistical and experimental errors are added in quadrature into the integration and displayed with corresponding error bars. 
{
The integrals from the JAM fits are shown as bands corresponding to $\pm$ 1-$\sigma$ CL.}

{
For our second investigation, we define a sequence of {
{ fixed}} bins in $x$, each
with a width of $\Delta x = 0.05$.
The measured $g_1$ points are
averaged within each of these $x$-bins for
each of the
same $Q^2$ bins as before, and the averages are plotted 
against
the nominal  $Q^2$ values. Again,
the JAM fits are treated in the same way and shown as bands
together with the data. In this case, we show results both for the resonance region and beyond
$W=2$~GeV, depending on the x-range.}


\section{Results}

\begin{figure*}[t]
\centering
\includegraphics[width=\textwidth]{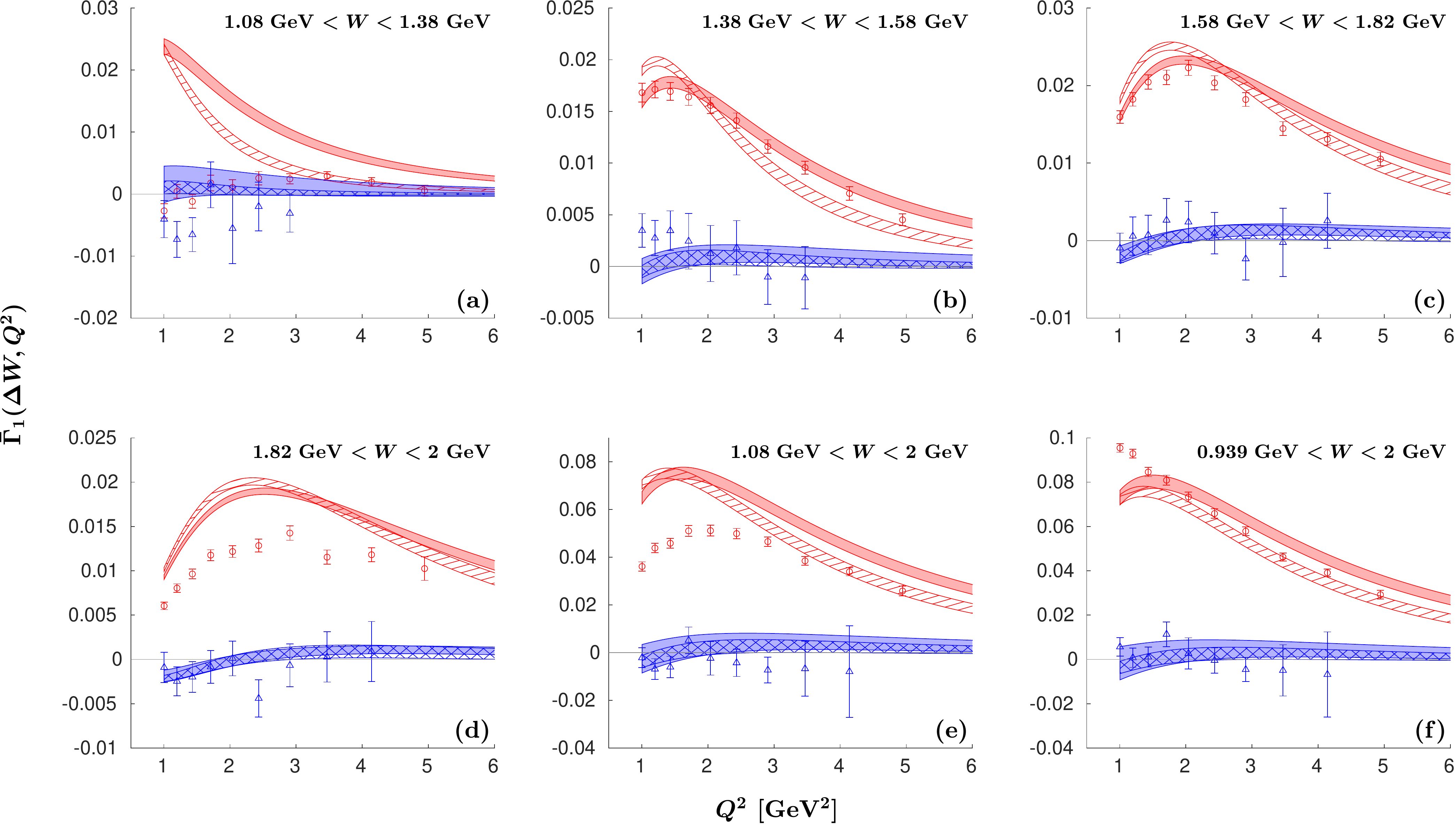}
\caption{
    [Color Online] Test of duality for truncated integrals. 
     We show integrals
     $\bar{\Gamma}_1(\Delta W, Q^2)$ of the spin structure
     function $g_1(x,Q^2)$ over regions in $x$ corresponding to six fixed regions $\Delta W$ of
     final state mass $W$, plotted as a function
     of $Q^2$.
     Panel (a): The region of the first excited state of the nucleon, the $\Delta(1232)$ resonance.
     Panel (b): The region of the $N(1440)1/2^+$, $N(1520)3/2^-$, and $N(1535)1/2^-$ resonances.
     Panel (c): The region including the $N(1680)5/2^+$ resonance.
     Panel (d): The remainder of the  customary resonance region, 1.82 GeV $< W <$ 2 GeV.
     Panel (e): The sum of regions (a) through (d), {\em i.e}, 
     the entire resonance region 1.07 GeV $< W <$ 2 GeV.
     Panel (f): Same as panel (e), with the elastic peak included: $0.938$~GeV~$ < W < 2$~GeV (corresponding
     to a range in $x$ extending all the way to $x = 1$).
     The top (red) bands 
     and data points (circles) are for the proton, and the bottom (blue) bands and data points (triangles) are for the neutron. The data points are shown with statistical
     and systematic uncertainties added in quadrature (error bars). The solid bands show the full prediction from the
     extrapolated JAM fit, including target mass and higher
     twist contributions. The striped band (proton) and the cross-hatched band (neutron) show the results including only the leading twist contribution.}
\label{fig:DualityPlots}
\end{figure*}

\begin{figure*}[t]
\centering
\includegraphics[width=\textwidth]{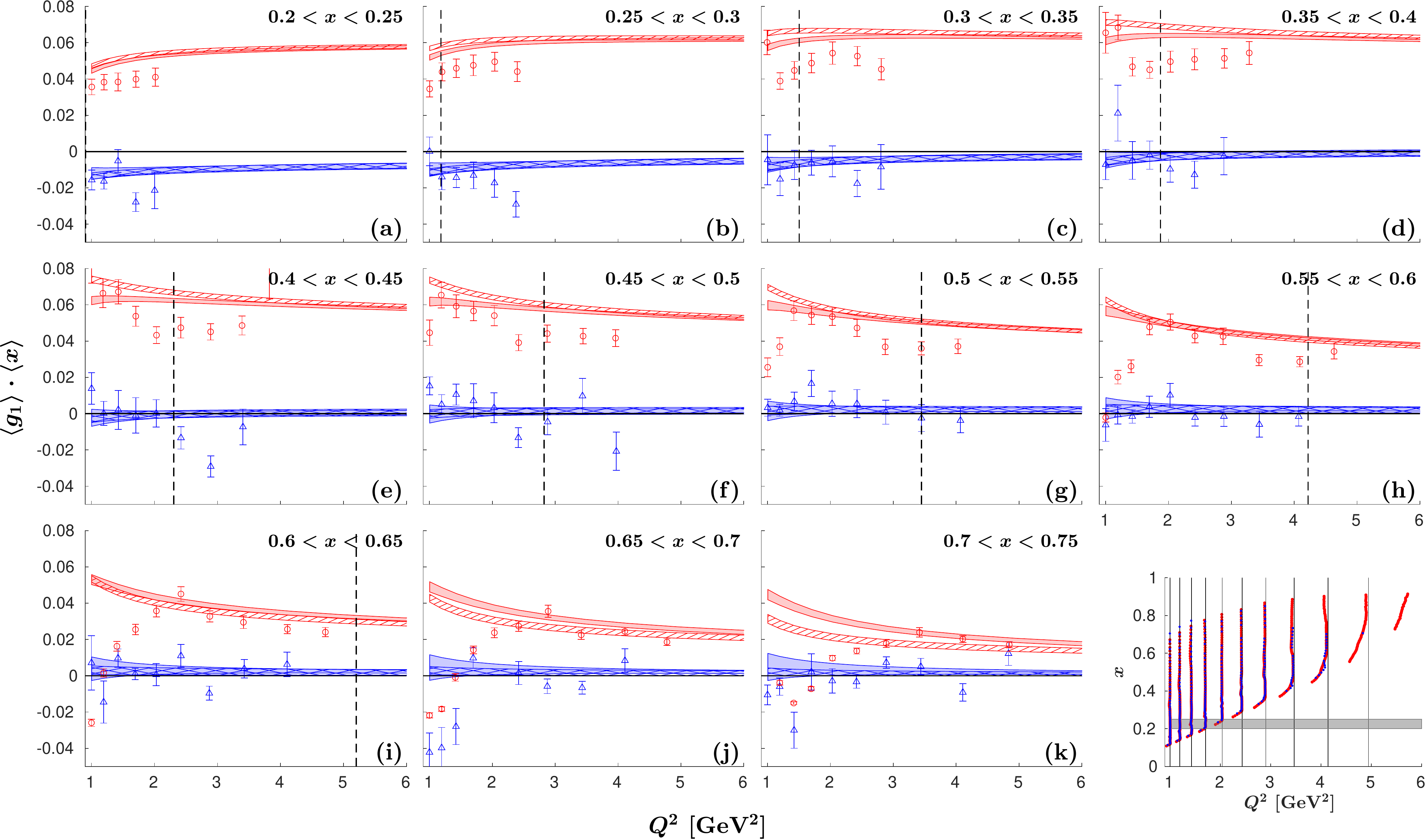}
    \caption{
     [Color Online]
     Approach towards the scaling limit for the structure function
     $g_1(x,Q^2)$, averaged over 11 different $x$-bins of width $\Delta x = 0.05$,
     as a function of $Q^2$.
    Data and JAM bands are shown multiplied with the average
     $x$ for each bin for better clarity; symbols are as in Fig.~\ref{fig:DualityPlots}.
     The vertical dashed line indicates the limit $W=2$ GeV of the resonance
     region, which lies to the left. 
     Last panel: kinematic location of all data points from EG1b in
     the $x$ {\it vs.} $Q^2$ plane (red - proton, blue - neutron). The grey band indicates a sample
     interval in $x$ over which the data are averaged (corresponding to panel (a)), and the  
     vertical lines indicate the nominal central values of each 
     $Q^2$ bin.}\label{fig:ScalingPlots}
\end{figure*}

{
In this Section, we  present the results of
our two tests of duality. We begin by showing the truncated integrals $\bar{\Gamma}_1(\Delta W, Q^2)$
for both protons and neutrons
over each of our  six different ranges in $W$ in Fig.~\ref{fig:DualityPlots}. The first four panels (a-d) 
in this figure test local duality in each of the four ranges of prominent nucleon resonances, while
the last two panels (e and f) test global duality over the entire resonance region. 

The first panel (a) shows the truncated integrals over the region of the 
of the lowest-lying Delta resonance. It is clear that duality does not work
very well in this region, especially for the proton (upper bands and data points). 
Both the proton and the neutron data are either negative (neutron, lower band and data points) or
close to zero (proton), while the PDF extrapolation for both
is positive (significantly so for the proton).
This disagreement is due to the well-known
fact that the excitation of the Delta resonance is dominated
by a $M1$ transition, for which the final state helicity
$3/2$ has a stronger coupling than the final state helicity
$1/2$, leading to a negative (virtual)
photon asymmetry $A_1$ and, in consequence, a negative
value for $g_1$. Meanwhile, the relatively low value of $W$
corresponds to large values of $x$, where most PDF fits 
predict a rising positive asymmetry. The JAM fit furthermore
predicts a rather strong positive contribution from finite target mass
and higher twist effects (solid upper band), making the disagreements
more pronounced.
Convergence of the resonance region data towards the PDF extrapolation only
begins around $Q^2 > 3$ GeV$^2$ for the neutron, and even later for the
proton if non-leading twist is included in the extrapolation.
Consequently, local duality is not a good assumption for spin structure
functions in the Delta region. 

The next two resonance regions (Panels b and c) show remarkably good agreement between
the data and the extrapolated PDF bands
(in particular the extrapolations including
higher twist), indicating that
``local duality'' works well for these resonances. It may 
therefore be possible to include truncated integrals over these
two regions in future PDF fits which include higher twist contributions,
helping to constrain these fits at high $x$ where experimental data
are scarce.
The remaining region, up to $W = 2$ GeV, shows again
a deviation of the data which tend to lie below the
PDF fits. Once again, this is consistent with the 
assumption that this region has a strong contribution from
various Delta-resonances, where the helicity-3/2 contribution
dominates at small $Q^2$.

Finally, for a test of ``global duality'', we integrate the data
over the entire resonance region, $1.072$ GeV $ < W < 2$ GeV (panel e).
We see that the data for the proton fall
short of the extrapolated PDF results up to rather high $Q^2 > 3.5$ 
GeV$^2$, and even higher for the band including higher twist. 
The neutron data have larger uncertainties, but also tend to lie consistently
below the extrapolated PDF results. This finding indicates that the
very slow approach towards duality in the (2) Delta resonance region(s) spoils
global duality, in contrast to the case of unpolarized structure functions of the proton.
However, this picture changes drastically if the integral is extended all the way to $x=1$,
including the elastic peak in the data ($0.938$ GeV $ < W < 2$ GeV, panel f).
It is remarkable how the negative deviations in the
lowest and highest $W$ regions (both
populated by Delta resonances) are compensated by the
inclusion of the elastic peak to get a rather rapid
approach to ``global duality''.
For the proton a clear (and non-trivial)
agreement between data and PDF prediction is observed, starting around
$Q^2 = 1.4$ GeV$^2$.
For the neutron the predictions from
the PDF fits as well as the data are mostly consistent
with zero. There may be a slight tendency for
the data to fall below zero at high $Q^2$,
which would agree with the observation that the $d-$quark
polarization appears to remain negative up to the highest
$x$ measured so far~\cite{Zheng:2004ce}. Overall, 
the integral over the entire resonance region can provide another constraint
for future polarized PDF fits, {\em provided} the elastic peak is included 
in the integral.
}

In our second analysis, we are averaging the EG1b data and JAM
PDF fits over fixed intervals in $x$ for each of our $Q^2$ bins,
to study the approach towards scaling for this averaged structure function
$g_1(x,Q^2)$. The data and JAM predictions are integrated over bins of width $\Delta x = 0.05$
and then divided by $\Delta x$
to obtain the average $\langle g_1 \rangle$. They are then multiplied by the bin centroid in $x$ for better visibility - see Fig.~\ref{fig:ScalingPlots}. 
In contrast to the previous analysis, we include in these figures {\em all} data from
EG1b, from both the resonance and the DIS region, with the
boundary between the two indicated by the vertical dashed line at 
$Q^2=(W_{\rm limit}^2- m^2)/(1/x-1)$, with $W_{\rm limit} = 2$ GeV. 
{
The last panel in Fig.~\ref{fig:ScalingPlots}
shows all of the EG1b data points, 
with the grey band indicating the range of data points integrated over for 
a sample $x$-bin.
For the first  $x$-bin (panel (a) of Fig.~\ref{fig:ScalingPlots}), 
all data points are already above the resonance region
($W > 2$ GeV) and show reasonably flat $Q^2$-dependence, albeit slightly below
the fit to all world data. 

For the next seven $x$-bins (panels), some non-trivial structure can be seen just 
to the left of the boundary at $W = 2$ GeV, while the data at both higher and
lower $Q^2$ (but above the $\Delta(1232)$ resonance region) seem to agree with the $Q^2$-behavior predicted by the extrapolated
pQCD fit. These ``dips'' occur in the higher-lying resonance region where
we already observed a slow convergence to the scaling limit, due to the presence
of some $\Delta$ resonances. Finally, at the highest $x$-bins in our sample (bottom row),
the data seem to converge more quickly towards the extrapolated pQCD fit, even
far away from the $W=2$ GeV limit (for the last 2 panels, {\em all} data are in
the resonance region). In this higher $x$ region, an approach to scaling is observed above
$Q^2=3$ GeV$^2$, basically as soon as $W$ is safely above the region of the 
$\Delta(1232)$ region.
Thus it appears as if the approach
to scaling may set on early at larger $x$ values, which would be very beneficial for the goal of extracting the behavior of
spin structure functions at large $x$, a topic of continuing high 
interest~\cite{Liu:2019vsn}. 
For tables for all $x$ ranges
see the Supplemental Material at [URL will be inserted by publisher]. }

\section{Conclusion}

In this paper, we present the most detailed study of quark-hadron duality in
the spin structure function $g_1$ to date, for both the proton and, 
for the first time, the neutron. We study several different formulations of duality,
and find that duality seems to hold much better (at smaller momentum
transfer) in some cases than in others. In particular, we conclude
the following:
\begin{itemize}
    \item When forming integrals over kinematic regions 
    corresponding to specific resonance peaks, we observe good
    agreement between the measured data and the extrapolation from
    pQCD fits whenever
     several resonances with different
    spins contribute, {\it i.e.} the for the regions $W = 1.38$ GeV...1.58 GeV 
    (including the N$(1440)1/2^+$, N$(1520)3/2^-$, and 
N$(1535)1/2^-$ resonances) and $W = 1.58$ GeV...1.82 GeV (several
higher-lying resonances). In contrast, in the region dominated by the ground-state Delta-resonance ($W =$ 1.08 GeV...1.38 GeV) and the region $W = 1.82$ GeV...2 GeV with several Delta resonances, we
observe a much slower approach of the measured integrals towards
the extrapolated PDF fits with $Q^2$. This is likely due to the fact
that, at least at moderately low $Q^2$, for the excitation of Delta resonances
the transition to final-state helicity 3/2 dominates.
\item If we integrate over the entire resonance region up to $W=2$,
including the elastic peak at $W = 0.938$ GeV, a rather rapid convergence
towards the extrapolated PDF fits is observed: Global duality seems to work
in spin structure functions. However, excluding the elastic peak leads to
rather slow convergence of the truncated integral to the extrapolated pQCD
expectation, due to the outsized influence of the negative contribution
from the $\Delta(1232)$
\item If instead we integrate over fixed bins in $x$, with different
resonances contributing at different $Q^2$, we find that for lower
values of $x$, the transition with $Q^2$ towards a smooth scaling curve occurs 
only if the value of $Q^2$ is high enough so that $W > 2$ GeV. Conversely,
for the highest $x$ values, we observe that the approach towards a smooth
scaling curve (and the extrapolated PDF fits) occurs even below $W = 2$ GeV,
albeit at a higher $Q^2 \approx 3$~GeV$^2$. This may be due to the fact that at higher $Q^2$,
resonant and non-resonant contributions with different asymmetries average
out, leading to a ``precocious'' approach to scaling (or a different
form of local duality). This observation supports the idea that, for high
enough $x$ and $Q^2$, even data in the resonance region may be used
to constrain (polarized) parton distribution functions. 
Being able to include data in the resonance
region and {\it a fortiori} at moderate $W^2 < 10$ GeV$^2$ - a 
limit often imposed on PDF fits - will help with the goal
to pin down more precisely the quark polarization of both
types of valence quarks in the limit $x \rightarrow 1$, which
is still an open question at this time.
\end{itemize}

\acknowledgments{}
We would like to thank our collaborators on CLAS experiment EG1b 
and the members of the JAM collaboration for their help with the
analysis presented. We are particularly indebted to W. Melnitchouk for his valuable comments and to
J. Ethier, who provided us with the extrapolated JAM results.
This work was supported by the Department of Energy under Contract
DE-FG02-96ER40960 (ODU). The work of N.S. was supported by the DOE, Office of Science, Office of Nuclear Physics in the Early Career Program. We would like to dedicate this
paper to the memory of Dr. Robert Fersch, who passed away
in the summer of 2022. 


%

\end{document}